\documentclass[prd,aps,preprint,preprintnumbers]{revtex4-1}

\usepackage{epsfig,subfigure}
\usepackage{amsmath,amsfonts}

\usepackage{epstopdf}

\def\as{\ensuremath{\alpha_{s}}}
\def\a0{\alpha_0}

\def\vep{\varepsilon}

\def \ep{\epsilon}

\def\bea {\begin{eqnarray}}
\def\eea {\end{eqnarray}}

\def\be {\begin{equation}}
\def\ee {\end{equation}}
\def\bi {\begin{itemize}}
\def\ei {\end{itemize}}

\begin{document}

\preprint{YITP-SB-11-33}

\renewcommand{\thefigure}{\arabic{figure}}

\title{Gauge theory webs and surfaces}

\author{Ozan Erdo\u{g}an, George Sterman}

\affiliation{C.N.\ Yang Institute for Theoretical Physics and Department of Physics and Astronomy\\
Stony Brook University, Stony Brook, New York\ 11794-3840, USA}

\date{\today}

\begin{abstract}
We analyze the perturbative cusp and closed polygons of Wilson lines for
massless gauge theories in coordinate space, and express them
as exponentials of  two-dimensional integrals.  
These integrals have geometric
interpretations, which link renormalization scales with invariant
distances.     
\end{abstract}

\maketitle

\section{Introduction}

Gauge field path-ordered exponentials \cite{Bialynicki-Birula,Yang:1974kj,Wilson:1974sk} or Wilson lines, 
 represent the interaction of energetic partons with relatively softer radiation
in gauge theories.    
For constant velocities, ordered exponentials of semi-infinite length
correspond to the eikonal approximation for energetic partons.
Classic phenomenological applications of ordered exponentials include soft radiation limits in
deeply inelastic scattering \cite{Korchemsky:1993uz} and parton pair
production and electroweak annihilation 
\cite{Korchemsky:1994is,Belitsky:1998tc,Kelley:2011ng}\@. They
appear as well  in the treatment of parton
distributions~\cite{Laenen:2000ij,Cherednikov:2013bxa}\@. 
In all these cases, the electroweak current is represented by a color
singlet vertex at which lines in the same color representation but
with different velocities are coupled. This vertex is often referred
to as a cusp.   

Cusps also appear as vertices in polygons formed from Wilson lines \cite{Korchemskaya:1992je}, which have been studied extensively in the context of their duality to scattering amplitudes in ${\cal N}=4$ SYM theory~\cite{Drummond:2007aua,Alday:2007hr,Alday:2008yw,Chien:2011wz,Basso:2013vsa}\@. In 
the strong-coupling limit of this theory, gauge-gravity duality relates the cusp
and polygons to the exponentials of two-dimensional surface
integrals. Surfaces bounded by open and closed paths of ordered exponentials are also a classic ingredient in lattice 
\cite{Wilson:1974sk} and large-$N_c$ \cite{'tHooft:1973jz} paradigms for confinement in quantum chromodynamics.

In this paper, we show that in any gauge theory with massless vector bosons the cusp matrix
element for lightlike Wilson lines can be expressed as the exponential
of an integral over a two-dimensional surface, a result with
applications as well to polygons formed from ordered exponentials. The
corresponding integrand is an infrared finite function of the gauge theory coupling,
evaluated for each point on the surface at a scale given by the
invariant distance from that point  to the cusp vertex. This result
extends to all orders in perturbation theory.   

The set of all virtual corrections for the cusp \cite{Korchemsky:1987wg} is formally identical to a vacuum expectation value, and can be
written as
\bea
\Gamma^{(f)}(\beta_1,\beta_2)
=
\bigg \langle 0\left| T\bigg(  \Phi^{(f)}_{\beta _2}(\infty,0)\, \Phi^{(f)}_{\beta _1}(0,-\infty) \bigg)\right|0  \bigg \rangle\, ,
\label{eq:Gammadef}
\eea
in terms of constant-velocity ordered exponentials,
\bea
\Phi^{(f)}_{\beta _i}(x+\lambda\beta_i,x)
&\ &
=
{\cal P}\exp \left (-ig\int_0^\lambda d\lambda'\beta _i\cdot A^{(f)}(x+\lambda'\beta_i)
\right )\, .
\label{eq:Phidef}
\eea
Here  $f$ labels a representation of the gauge group 
 and $\beta_i$ is a four-velocity,
taken lightlike in the following.  
The combination of ordered exponentials in Eq.~(\ref{eq:Gammadef})
corresponds to a partonic
process with spacelike momentum transfer. For correspondence to a
timelike process like pair creation, $\Phi^{(f)}_{\beta_1}(0,-\infty)$
can be replaced by 
$\Phi^{(\bar  f)}_{\beta_1}(\infty,0)$\@. 
The imaginary parts for timelike configurations have been discussed recently in Ref.\ \cite{Laenen:2014jga}\@. 
Corrections to partonic scattering~\cite{Kidonakis:1998nf,Bauer:2001yt,Mitov:2009sv,Beneke:2010da,Ferroglia:2009ii,Kidonakis:2010dk,Gardi:2010rn,Kelley:2010fn,Jouttenus:2011wh,Gardi:2013ita,Gardi:2013jia}\@ involve the coupling of more than two
ordered exponentials at a
point~\cite{Brandt:1981kf,Korchemskaya:1994qp}. In this paper we study the all-orders properties of the single cusp and of polygons with sequential cusps, computed perturbatively in coordinate space.

Perturbative corrections to the unrenormalized cusp, Eq.~(\ref{eq:Gammadef})
are scaleless, and hence vanish in dimensional regularization. The
ultraviolet poles of (\ref{eq:Gammadef}) determine the anomalous
dimension of the cusp, and can be used to define a renormalized expansion, both for the cusp and for polygons formed from ordered exponentials
of finite length~\cite{Brandt:1981kf,Korchemskaya:1992je}\@. 
For the cusp in an asymptotically free theory, however, neither its ultraviolet nor its infrared behavior can be considered as truly physical.
At very short distances, dynamics is perturbative and recoil cannot be neglected.
At very long distances, dynamics is nonperturbative, and dominated by the hadronic
spectrum.   In this discussion,
we will regard the cusp as an interpolation between these asymptotic
regimes.   We will
concentrate on the structure of the integrals in the intermediate
region, although we also discuss their renormalization. 

We begin in Sec.~\ref{sec:exp} with a review of exponentiation for
products of ordered exponentials,
a result that extends to arbitrary products of such lines and to closed loops.   Section~\ref{sec:coord} recalls the
coordinate-space picture of exponentiation in terms of web diagrams  and introduces the cancellation of
subdivergences of webs. It is in this discussion that a surface  interpretation
of the exponent emerges. We provide a two-loop illustration of subdivergence
cancellation, motivate its generalization to all orders, and give an all-orders prescription for the calculation of the 
cusp exponent.
In Sec.~\ref{sec:poly}, we apply these ideas to
multi-cusp polygonal Wilson loops.

\section{Exponentiation and Momentum-Space Webs}
\label{sec:exp}

\begin{figure}[b]
\centering
\subfigure[]{\includegraphics[width=2cm]{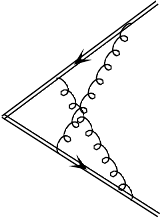}\label{fig:cross}}
\hspace{1cm}
\subfigure[]{\includegraphics[width=2cm]{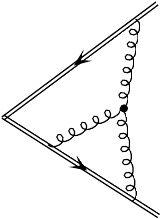}\label{fig:3g}}
\hspace{1cm}
\subfigure[]{\includegraphics[width=2cm]{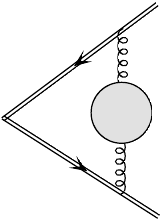}\label{fig:se}}
\hspace{1cm}
\subfigure[]{\includegraphics[width=2cm]{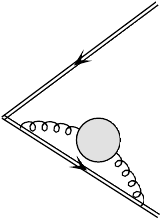}\label{fig:se2}}
\caption{Two-loop web diagrams, referred to in the text as:
  \subref{fig:cross}~$E_{\rm cross}$, \subref{fig:3g}~$E_{3g}$,
  \subref{fig:se}--\subref{fig:se2}~$E_{\rm se}$\@. Web
  diagram~\subref{fig:cross} has the modified color factor,
  $C_aC_A/2$, where $a$ refers to the representation of the Wilson
  lines. For diagram~\subref{fig:cross}, the web
color factor differs from its original color factor, while all other
color factors are the same as in the normal expansion. Diagrams related by
top-bottom reflection are not shown.}
\label{webfig}
\end{figure}

The cusp has long been known \cite{Gatheral:1983cz} to be 
 the exponential of a sum of special diagrams called webs, which are  irreducible by cutting two
eikonal lines.
 We represent this result as 
\bea
\Gamma(\beta_1,\beta_2,\vep)\
=\
\exp E(\beta_1,\beta_2,\vep)\, ,
\eea
in $D=4-2\vep$ dimensions.
The exponent $E$ equals a sum over web diagrams, $d$,
each given by a   group factor
multiplied by a diagrammatic integral, 
\bea
E(\beta_1,\beta_2,\vep)\ =\ \sum_{webs\ d} \overline{C}_d\; {\cal F}_d(\beta_1,\beta_2,\vep)\, ,
\eea
where ${\cal F}_d$ represents the momentum- or coordinate-space
integral for diagram~$d$\@. The coefficients of these integrals,
$\overline{C}_d$ are modified color factors.  Two-loop examples are
shown in Fig.~\ref{webfig}\@.

In momentum space we
can write the exponent $E$ as the integral over 
a single, overall loop momentum  that
runs through the web and the cusp vertex, assuming that
all loop integrals internal to the web 
have already been carried out.  The web is defined to  include  the necessary
counterterms of the gauge theory
\cite{Dotsenko:1979wb,Brandt:1981kf,Laenen:2000ij,Berger:2003zh}\@. Taking
into account the boost invariance of the cusp for massless loop
velocities, and the invariance of the ordered exponentials under rescalings of
the velocities $\beta_i$, we have for the
exponent the form,
\bea
E (\beta_1,\beta_2,\vep) \ =\ 
 \int \frac{d^D k}{(2\pi)^D}\, \frac{\beta_1\cdot\beta_2}{k\cdot \beta_1\, k\cdot \beta_2}\ \frac{1}{k^2}\  \bar w \left(\frac{k^2}{\mu^2},{k\cdot \beta_1\, k\cdot \beta_2\over \mu^2
\beta_1\cdot\beta_2},\alpha_s(\mu^2,\vep),\varepsilon\right)\, .
\label{eq:webexpk}
\eea
In addition, the webs themselves are renormalization-scale independent,
\bea
\mu{d\over d\mu}\; \bar w \left(\frac{k^2}{\mu^2},{k\cdot \beta_1\, k\cdot \beta_2\over \mu^2
\beta_1\cdot\beta_2},\alpha_s(\mu^2,\vep),\varepsilon\right)
=0\, .
\eea
This renormalization-scale invariance allows us to choose $\mu^2$ equal to either of
the kinematic arguments in the web.   A further important property of webs is the absence of collinear and soft subdivergences in the sum of all web diagrams.    
That is, in Eq.\ (\ref{eq:webexpk}), collinear poles are generated only when $k^2$ and either $k\cdot \beta_1$
or $k\cdot \beta_2$ vanish, 
infrared poles only when all three vanish and the overall
ultraviolet poles only when all components of $k$
diverge. Equation~(\ref{eq:webexpk}) thus organizes the same double poles
found in the corresponding partonic form
factors~\cite{Berger:2003zh,Magnea:1990zb,Catani:1998bh}\@. Arguments for these properties in
momentum space are given in 
Ref.~\cite{Berger:2003zh}, based on the factorization of soft 
gluons from fast-moving collinear partons.    These considerations suggest that 
when embedded in an on-shell amplitude or cross section, the web acts
as a unit, almost like a single gluon, dressed by arbitrary orders in
the coupling. In the following, we observe that this analogy can be extended
to coordinate space.

The form given above, in terms of webs, is for the unrenormalized
cusp. When renormalized by the minimal subtraction of ultraviolet
poles, the exponent $E$ can be written in the
form~\cite{Dixon:2008gr}, 
\bea
 \hspace{-3mm} E_{\rm ren}(\alpha_s(\mu^2),\vep)\ =\ - \frac{1}{2} 
  \int_0^{\mu^2} \frac{d \xi^2}{\xi^2} \left[
   \Gamma_{\rm cusp} \Big( \alpha_s
  \left(\xi^2 \right) \Big) \, \log \left(\frac{ \mu^2}{\xi^2}\right) \ -\ G_{\rm eik}(\alpha_s\left(\xi^2)\right)  \right] 
  \, ,
  \label{eq:E-ren}
  \eea
where $\mu^2$ is the renormalization scale, and where, here and below,
we have set $\beta_1\cdot\beta_2=1$.  At order $\as^n$, the leading
pole behavior of this exponent is proportional to $\Gamma_{\rm
  cusp}^{(1)}\,\as^n (1/\vep)^{n+1}$, with $\Gamma_{\rm
  cusp}^{(1)}(\as/\pi)$ 
the one-loop cusp anomalous dimension. Nonleading poles are generated
from higher orders in $\Gamma_{\rm cusp}$, from  $G_{\rm eik}$,
and from the $\vep$-dependence of the running  coupling in $D$
dimensions~\cite{Magnea:1990zb}\@.   
After renormalization in this manner, the cusp is a sum of infrared
poles in one-to-one correspondence with the ultraviolet poles that are
subtracted. The cusp anomalous dimension is given to two loops by
\bea
\Gamma_{\rm cusp,\, a}\ &=&\ \left(\frac{\alpha_s}{\pi}\right) C_a\left[1+\left(\frac{\alpha_s}{\pi}\right)K\right]\, ,
\nonumber\\
K\ &=&\ \left(\frac{67}{36}-\frac{\pi^2}{12}\right)C_A\,-\,\frac{5}{18}n_fT_f\, ,
\label{eq:Gamma2loop}
\eea
with $C_a=4/3,3$ for $a=q,g$ for QCD, $n_f$ the number of fermion flavors, 
and $T_f=1/2$\@. At one loop, $G_{\rm eik}$ is zero, and we will derive its two-loop form below.   Equation (\ref{eq:E-ren}) gives all the poles of the cusp, when reexpanded in terms of the coupling at any fixed scale.   We note that for timelike kinematics, the renormalization scale $\mu^2$ should be chosen negative \cite{Dixon:2008gr}\@.  

\section{Webs and Surfaces in Coordinate Space}
\label{sec:coord}

\subsection{The unrenormalized exponent and its surface interpretation}

The coordinate-space analog of Eq.\ (\ref{eq:webexpk})
is a double integral over two parameters, $\sigma$ and $\lambda$ that measure distances along 
the Wilson lines $\beta_1$ and $\beta_2$, 
respectively,
with a new web function, $w$, which
depends on these variables through the only
available dimensionless combination, $\lambda\sigma\mu^2$,
\bea
E
&=&
\int_0^\infty \frac{d\lambda}{\lambda}\, 
\int_0^\infty \frac{d\sigma}{\sigma}\ w (\alpha_s(\mu^2,\vep),\lambda\sigma\mu^2,\varepsilon)\, .
\label{eq:web2}
\eea
Here and below, we choose timelike kinematics. 
We emphasize that we are interested primarily in the form and symmetries of the
integrand, rather than its convergence
properties.    Nevertheless,
to separate infrared and ultraviolet poles
in the integration, it is necessary that the integrand, $w$ in Eq.\ (\ref{eq:web2})
be free of both infrared and ultraviolet divergences at $\varepsilon=0$ in 
renormalized perturbation theory (aside from the renormalization of the cusp itself).
As we shall see below,  Eq.\ (\ref{eq:web2}) with a finite web function 
leads to a renormalized cusp that is fully consistent with the momentum-space
form, Eq.\ (\ref{eq:E-ren}).   In this construction, all $\vep$ poles of the
exponent, and therefore the cusp, are then associated with the
integrals over $\lambda$ and $\sigma$ in~(\ref{eq:web2})\@.  

A direct, coordinate-space demonstration of the finiteness of the web
function is interesting in its own right, 
and is given in Ref.~\cite{Erdogan:2014gha}\@. Formally, such a
demonstration is necessary to extend  the proof of renormalizability
for cusps connecting massive lines \cite{Brandt:1981kf} to the massless case
\cite{Korchemskaya:1992je}\@. Here, we simply mention the essential
ingredients of such an argument.

Diagram by diagram, one may use the analytic  structure of the
coordinate integrations~\cite{Date:1982un} 
combined with a coordinate-space power-counting technique  to identify
the most general singular subregions in coordinate
space~\cite{Erdogan:2013bga}\@.    
In coordinate space, nonlocal ultraviolet subdivergences arise when a subset of vertices line up at finite distances
from the cusp along either of the lightlike Wilson lines, while other,
``soft'' vertices remain at finite distances.  
Such subdiagrams factorize, however, in much the same manner as in
momentum space~\cite{Bodwin:1984hc,Collinsbook}\@.    
Once in factorized form, combinatoric arguments show that divergent
integrals  cancel when all web diagrams are combined at a given
order~\cite{Erdogan:2014gha} in coordinate space, in much the same way as
in the momentum-space treatment of Ref.\ \cite{Berger:2003zh}\@.   
Finally, taking $\lambda$ and $\sigma$ as the positions of vertices
in the web diagrams furthest from the cusp,
there are no soft (infinite wavelength) divergences from integrations
over the internal vertices of webs in coordinate
representation, as shown in Ref.~\cite{Erdogan:2013bga}\@.

\begin{figure}[t]
\centering
\subfigure[]{\includegraphics[width=8.5cm]{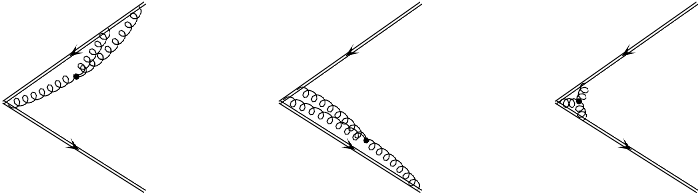}\label{fig:2lw1}}\\ 
\subfigure[]{\includegraphics[width=8.5cm]{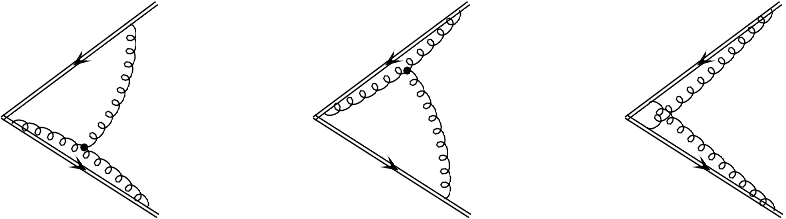}\label{fig:2lw2}}
\caption{Representation of singular regions for a two-loop web
  diagram.  \subref{fig:2lw1}~Single-scale regions, characteristic of webs.
  \subref{fig:2lw2}~Multiple-scale regions, associated with subdivergences that cancel in the
  sum of web diagrams.}
\label{fig:2loopwebs}
\end{figure}

As we shall 
shall illustrate below, it is possible to implement the cancellation of subdivergences at fixed positions, $\lambda$ and $\sigma$, along the ordered paths, specified by the vertices furthest from the cusp.   
Once this is done and the subdivergences thereby eliminated,
the integrals over all vertices of the web diagrams converge on scales
set by $\lambda$ and $\sigma$  in (\ref{eq:web2}), and the web acts as a unit.  
Singular behavior of the cusp arises as 
$\lambda$ and/or $\sigma$ vanish, and in these limits all web
vertices approach the directions of $\beta_1$ or $\beta_2$ together,
as in Fig.~\ref{fig:2lw1}\@. This is the 
perturbative
realization
of the web as a geometrical 
object.
Subdivergent configurations that cancel are illustrated in
Fig.~\ref{fig:2lw2}\@.

The web function $w$ constructed this way is again a renormalization group invariant, so that
in (\ref{eq:web2}), we may shift the renormalization scale to the product $(\lambda\sigma)^{-1}$, which
results in an expression with the coupling running as
the leading vertices move up and down the Wilson lines,
\bea
E
&=&
\int_0^\infty \frac{d\lambda}{\lambda}\, 
\int_0^\infty \frac{d\sigma}{\sigma}\, w\left(\alpha_s\left(1/\lambda\sigma\right),\varepsilon\right)\, .
\label{eq:web3}
\eea
In this all-orders form, dependence on the product $\lambda\sigma$ is entirely
through the running coupling, aside from the overall dimensional factor.
For ${\cal N}=4$ SYM theory, Eq.\ (\ref{eq:web3}) for the cusp holds as well 
at strong coupling \cite{Kruczenski:2002fb,Alday:2008yw,Alday:2007hr},
where the coordinates $\lambda$ and $\sigma$  also parametrize a surface.
The generality of these results
can be traced to the symmetries of the problem \cite{Kruczenski:2002fb}.
It is interesting to note, however, that in the strong-coupling
analysis, the product of internal coordinates $\lambda\sigma$, which
serves as the renormalization scale in Eq.\ (\ref{eq:web3}),  relates
the plane of the Wilson lines to a minimal surface in five dimensions.

\subsection{Web renormalization in coordinate space}

To derive a renormalized exponent  for the cusp in coordinate space, we will find it useful to
 expand the unrenormalized web function in (\ref{eq:web3}) in explicit powers of $\vep$, 
\bea
E(\vep)\ &=&\ \sum_{n=0}^\infty\ \vep^n\  \int_0^\infty \frac{d\lambda}{\lambda}\, 
\int_0^\infty \frac{d\sigma}{\sigma}\, w_n\left(\alpha_s\left(1/\lambda\sigma\right)\right)\, ,
\label{eq:w-vep-expand}
\eea 
where $w_n$ is the coefficient of $\vep^n$, noting that the coupling retains implicit $\vep$-dependence.   As noted above, the
renormalized exponent is determined by the ultraviolet poles of
these scaleless integrals. With this in mind, consistency with momentum-space pole structure in Eq.\ (\ref{eq:E-ren}) then clearly requires 
\be
w_0\left(\alpha_s\left(1/\lambda\sigma\right)\right)\ =\ 
-\  \frac{1}{2}\; \Gamma_{\rm cusp} \big(\alpha_s\left(1/\lambda\sigma\right)\big)
 \  .
\label{eq:web4}
\ee
For finite values of $\lambda$ and $\sigma$, only $w_0$ contributes to the unrenormalized integral in the $\vep\to 0$ limit.  To determine the renormalized cusp integral, however, we must take into account contributions from the boundaries $\lambda=0$ and $\sigma=0$, which produce poles that can compensate explicit powers of $\vep$ in Eq.\ (\ref{eq:w-vep-expand}).    
Such boundary contributions from terms $\vep^n w_n$ with $n>0$ in Eq.\ (\ref {eq:w-vep-expand}) generate  the anomalous dimension $G_{\rm eik}$ in the renormalized form, Eq.\ (\ref{eq:E-ren}).   

To compute $G_{\rm eik}$, we recall that the running coupling $\as(1/\lambda\sigma)$ remains a function of $\vep$
when reexpanded in terms of the coupling at any fixed scale, $\mu$, which we represent as
\bea
\alpha_s(1/\lambda\sigma)\ &=&\ \alpha_s(\mu^2)\, (\mu^2\lambda\sigma)^\vep\, \left( 1\ +\ \frac{\as(\mu^2)}{4\pi}\, \frac{b_0}{\vep}\, \left[ (\mu^2\lambda\sigma)^\vep - 1\right]   \ + \dots \right)
\nonumber\\
&\equiv& \bar{\alpha}_s \left(\as(\mu^2),(\mu^2\lambda\sigma)^\vep,\vep\right)\, ,
\label{eq:alpha-ls-mu}
\eea
where we exhibit only the dependence to order $\as^2$, which is all we
need here, and where $b_0=(11/3)C_A-(4/3)n_fT_f$\@. The subleading anomalous dimension $G_{\rm eik}$ is found from single poles in $E(\vep)$ after the $\lambda$ and $\sigma$ integrations.   These can arise at any order by combinations of an overall factor $\vep^n$ in (\ref{eq:w-vep-expand}) with poles in the expansion of the coupling, (\ref{eq:alpha-ls-mu}).   To identify such terms, we may conveniently take $\sigma<\lambda$  and multiply by 2, and reexpand $\as(1/\lambda\sigma)$ in terms of $\as(1/\lambda^2)$, schematically,
\bea
E(\vep) & = &  -\ \frac{1}{2} \int_0^\infty \frac{d\lambda}{\lambda} 
\int_0^\infty \frac{d\sigma}{\sigma}\ \Gamma_{\rm cusp}
\big(\alpha_s\left(1/\lambda\sigma\right)\big) \\
 &\  &  \hspace{10mm} +\ 2\; \sum_{n=1}^\infty\, \vep^n\,  \int_0^\infty
 \frac{d\lambda}{\lambda} 
\int_0^\lambda \frac{d\sigma}{\sigma}\  w_n\left( \bar
  \alpha_s\left(\as(1/\lambda^2),
    (\sigma/\lambda)^\vep,\vep\right)\right)\, . \nonumber
\eea
The renormalized exponent is defined as the remainder when all ultraviolet poles are subtracted  minimally at an arbitrary, fixed scale $\mu$.   Leading and nonleading poles are then generated by
\be
E_{\rm ren}(\vep,\as(\mu^2))\, =\,  - \frac{1}{2} \int_{1/\mu}^\infty \frac{d\lambda}{\lambda} 
\int_{1/\mu}^\infty \frac{d\sigma}{\sigma}\, \Gamma_{\rm cusp} \big(\alpha_s\left(1/\lambda\sigma\right)\big)
\, +\, \int_{1/\mu}^\infty \frac{d\lambda}{\lambda}\, G_{\rm eik}\left(\as(1/\lambda^2)\right)\, ,
\label{eq:E-ren-coord}
\ee
where the integrals are now defined by infrared regularization
($\vep<0$).   Simple changes of variables transform this expression
into the renormalized cusp momentum-space integrals given in
Eq.~(\ref{eq:E-ren})\@. 

\subsection{Lowest orders}

The lowest order expression for Eq.\ (\ref{eq:web2}) already illustrates the nontrivial
relationship between the renormalization scale and the positions of the 
vertices.
It is found directly from the coordinate-space gluon propagator in
Feynman gauge, 
\bea
D^{\mu\nu}(x^2)\ & = &\ \int \frac{d^Dk}{(2\pi)^D}\ e^{-ik\cdot x}\ \frac{-i\, g^{\mu\nu}}{k^2\ +\ i\epsilon}
\nonumber\\
&=&\ 
\frac{\Gamma(1-\vep)}{4\pi^{2-\vep}}\, \frac{-g^{\mu\nu}}{\left(-x^2+i\ep\right)^{1-\vep}}\, .
\eea 
The resulting expression for the unrenormalized exponent is
\bea
E^{\rm (LO)}\, &=& \  -\ 
 C_F\, \frac{\Gamma(1-\vep)}{2} 
\int_0^\infty \frac{d\lambda}{\lambda}\, 
 \frac{d\sigma}{\sigma}\left(\frac{\alpha_s(\mu^2)}{\pi}\right) (2\pi
 \lambda\sigma \mu^2)^{\vep}\, , \\
 &=& 
 \  -\ 
 \frac{C_F}{2}\, \left(1 \ +\ \vep^2\, \frac{\pi^2}{12}\right)\;  
\int_0^\infty \frac{d\lambda}{\lambda}\, 
 \frac{d\sigma}{\sigma}\left(\frac{\alpha_s(\mu^2)}{\pi}\right) (2\pi
 e^{\gamma_E} \lambda\sigma \mu^2)^{\vep}\, , \nonumber
\eea
where in the second form we have expanded the integrand to order $\vep^2$.   The corresponding renormalized exponent is
\bea
 E_{\rm ren}^{\rm (LO)}\, &=& \ -\
 C_F\, \frac{\Gamma(1-\vep)}{2} 
\int_{1/\mu}^\infty \frac{d\lambda}{\lambda}\, \int_{1/\mu}^\infty
 \frac{d\sigma}{\sigma}\left(\frac{\alpha_s(\mu^2)}{\pi}\right) (2\pi \lambda\sigma \mu^2)^{\vep}\, ,
\
\label{eq:webLO}
\eea
which is precisely Eq.\ (\ref{eq:E-ren-coord}) to lowest order.   Here and below, for definiteness we choose the Wilson lines in fundamental representation.

At two loops, the diagrams of Fig.~\ref{webfig} can be used to illustrate both the cancellation of subdivergences in the sum of web diagrams, and the manner in which we identify the parameters $\lambda$ and $\sigma$, which together define the position of the web function.  Our calculations are carried out with ultraviolet regularization ($D<4$).   These coordinate-space integrals have appeared in the literature  before, of course, and the calculations we exhibit below are closely related to those of Refs.\ \cite{Korchemskaya:1992je} and \cite{Drummond:2007aua}, also carried out in dimensional regularization.   We present them again, however, in a form that shows explicitly how the cancellation of subdivergences occurs at fixed positions for the web along the lightlike paths, already in the unrenormalized forms.

The calculation of the crossed-ladder diagram, Fig.~\ref{fig:cross},
is particularly simple in coordinate space. It is just the integral of
two gluon propagators over the eikonal parameters, 
\be E_{\mbox{\scriptsize cross}}\ =\ \mathcal{N}_{\mbox{\scriptsize
    cross}}(\vep)
\int^{\infty}_0d\lambda_2\int^{\lambda_2}_0d\lambda_1\int^{\infty}_0d\sigma_2\int^{\sigma_2}_0d\sigma_1
\,\frac{1}{(2\lambda_2\sigma_1+i\epsilon)^{1-\vep}}\frac{1}{(2\lambda_1\sigma_2+i\epsilon)^{1-\vep}}
\ , \ee
where the prefactor is given by
\bea \mathcal{N}_{\mbox{\scriptsize
    cross}}(\vep) &\equiv& -
\left(\frac{\alpha_s}{\pi}\right)^2
C_AC_F\,\frac{\Gamma^2(1-\vep)}{2}\,(\pi\mu^2)^{2\vep}
\ . \eea
For the color factor in this web diagram, we keep only the $C_AC_F/2$ contribution, as mentioned above.  For $\vep>0$, we choose to integrate over the inner eikonal parameters, and identify $\lambda_2\equiv \lambda$ and $\sigma_2\equiv \sigma$ in the general form of Eq.\ (\ref{eq:web3}), giving
\bea
E_{\mbox{\scriptsize cross}} &=& -\left(\frac{\alpha_s}{\pi}\right)^2
 C_AC_F\,\frac{\Gamma^2(1-\vep)}{8\,\vep^2}\,(2\pi\mu^2)^{2\vep}  \int_0^\infty \frac{d\lambda\ d\sigma}{\hspace{4mm} (\lambda\sigma)^{1-2\vep}} \ . \eea
This expression has overall double ultraviolet poles in addition to two scaleless
(surface) integrals along the Wilson lines. The singular behavior of the coefficient arises from $\lambda_1\ll \lambda$ and $\sigma_1\ll \sigma$, a ``subdivergent" configuration, in which the two gluons approach different Wilson lines.   The contributions from these regions
will be canceled by corresponding terms from the three-gluon diagrams.

We now turn to the diagrams with a three-gluon coupling, one of which is shown in Fig.~\ref{fig:3g}, referred to below as $E_{3g}$.
In the expression for $E_{3g}$, we introduce upper limits, $\Lambda$ and $\Sigma$ on
the two paths. For the simple cusp, we will take the limit
$\Lambda,\, \Sigma \rightarrow \infty$.   We return to the finite case in the discussion of polygons. 

After evaluation of the three-gluon vertex, using $\beta_2^2=0$, $E_{3g}$ can be written as
\bea
E_{3g} &=&\ \mathcal{N}_{3g}(\vep)\int
d^Dx\int^{\Sigma}_0d\sigma\,\frac{1}{(-x^2+2\sigma
  x\cdot\beta_1+i\epsilon)^{1-\vep}} \nonumber
\\
 &\ & \hspace{3mm}\times\left[
  \int^{\Lambda}_0d\lambda_1\int^{\Lambda}_{\lambda_1}d\lambda_2\,\frac{1}{(-x^2+2\lambda_1x\cdot\beta_2+i\epsilon)^{1-\vep}}\frac{2x\cdot\beta_2(1-\vep)}{(-x^2+2\lambda_2x\cdot\beta_2+i\epsilon)^{2-\vep}}\right.
   \nonumber 
   \\
 &\ &  \left. \hspace{6mm}
 -\int^{\Lambda}_0d\lambda_2\int^{\lambda_2}_0d\lambda_1\,\frac{2x\cdot\beta_2(1-\vep)}{(-x^2+2\lambda_1x\cdot\beta_2+i\epsilon)^{2-\vep}}\frac{1}{(-x^2+2\lambda_2x\cdot\beta_2+i\epsilon)^{1-\vep}}\right] \nonumber  \\  
 &=&\  \mathcal{N}_{3g}(\vep)\int
d^Dx\int^{\Sigma}_0d\sigma\,\frac{1}{(-x^2+2\sigma
  x\cdot\beta_1+i\epsilon)^{1-\vep}}  \label{3v01}
\\
 &\ & \hspace{3mm}\times\left[
   \int^{\Lambda}_0d\lambda_2\, \frac{1}{(-x^2+2\lambda_2x\cdot\beta_2+i\epsilon)^{1-\vep}}\
   \int^{\lambda_2}_0d\lambda_1\,  \frac{\partial}{\partial\lambda_1} \left(\frac{1}{(-x^2+2\lambda_1x\cdot\beta_2+i\epsilon)^{1-\vep}}\right)
   \right. \nonumber 
   \\
 &\ &  \left. \hspace{6mm}
   - \int^{\Lambda}_0d\lambda_1\, \frac{1}{(-x^2+2\lambda_1x\cdot\beta_2+i\epsilon)^{1-\vep}}\
   \int^{\Lambda}_{\lambda_1}d\lambda_2\, \frac{\partial}{\partial\lambda_2} \left(\frac{1}{(-x^2+2\lambda_2x\cdot\beta_2+i\epsilon)^{1-\vep}}\right) \right]
 \, , \nonumber
 \eea
where in this case the numerical prefactor is
\bea \mathcal{N}_{3g}(\vep) & = & 
-i\left(\frac{\alpha_s}{\pi}\right)^2
C_AC_F\,\frac{\Gamma^3(1-\vep)}{8\pi^{2-\vep}}\,(\pi\mu^2)^{2\vep}
\ . \eea
In the second equality of Eq.\ (\ref{3v01}), we isolate two total derivatives, in the variables $\lambda_1$ and $\lambda_2$.
We shall carry out these two integrals first, at fixed values of the
other path parameters and of $x^\mu$.   

There is a suggestive way of interpreting the total derivatives in
Eq.~(\ref{3v01}), starting by recognizing that the ``propagator'' for
the Wilson line is a  step function, for example, 
$\theta(\lambda)$, with  ``equation of motion'' 
$\partial_\lambda\theta(\lambda)=\delta(\lambda)$\@. In these terms, the  $\lambda_1$ or
$\lambda_2$ integrals over total derivatives can also be thought of as
the result of integration by parts and the use of the equation of
motion.     In the term with $\partial/\partial_{\lambda_2}$, the equation of motion sets
$\lambda_2=\lambda_1$ and $\lambda_2=\Lambda$\@. As $\Lambda
\rightarrow \infty$ for fixed $x^\mu$, the  term with $\lambda_2=\Lambda$ vanishes 
as a power for any $\varepsilon<1/2$\@. The vanishing of such
contributions, through the cancellation of propagators, is an
ingredient in the gauge invariance of the cusp, which
generalizes to the gauge invariance of partonic amplitudes \cite{'tHooft:1971fh}.
We shall take the limit $\Lambda\rightarrow \infty$ first, at fixed values of the remaining integration variables after using the eikonal
equation of motion.   We will confirm below that this prescription gives a gauge-invariant result for the cusp after summing over diagrams.  We will 
evaluate the term from $\lambda_2 =\Lambda$, which by itself is gauge
dependent, in the Appendix.

\begin{figure}[b]
\centering
\subfigure[]{\includegraphics[width=2.2cm]{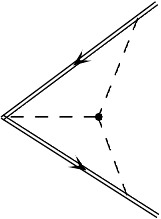}\label{fig:ses}}
\hspace{1cm}
\subfigure[]{\includegraphics[width=2.2cm]{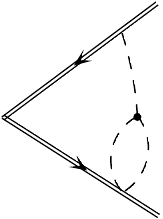}\label{fig:pse}}
\hspace{1cm}
\subfigure[]{\includegraphics[width=2.2cm]{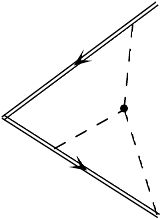}\label{fig:end}}

\caption{(a) 3-scalar diagram (b) Pseudo-self-energy diagram (c) End-point diagram.}
\label{fig:sespse}
\end{figure}

Returning to Eq.\  (\ref{3v01}), we  now integrate over the
total-derivative integrals, $\lambda_1$ in the first term and over
$\lambda_2$ in the second, and get
\bea
 E_{3g}\ &=&\ \mathcal{N}_{3g}(\vep)\int
d^Dx\int^{\Sigma}_0d\sigma\,\frac{1}{(-x^2+2\sigma
  x\cdot\beta_1+i\epsilon)^{1-\vep}} 
  \nonumber
\\
&\ & \hspace{5mm} \times\ \int^{\Lambda}_0 d\lambda \left[ -\, \frac{1}{(-x^2+i\epsilon)^{1-\vep}}\frac{1}{(-x^2+2\lambda
      x\cdot\beta_2+i\epsilon)^{1-\vep}}\right. 
  \,+\, \frac{2}{(-x^2+2\lambda x\cdot\beta_2+i\epsilon)^{2-2\vep}} 
 \nonumber \\
 &\ & \hspace{24.5mm} \left.
 -\, \frac{1}{(-x^2+2\Lambda
     x\cdot\beta_2+i\epsilon)^{1-\vep}}\frac{1}{(-x^2+2\lambda
     x\cdot\beta_2+i\epsilon)^{1-\vep}}\right] 
     \nonumber\\
     &\equiv&\ E_{3s}\ +\  2\,E_{\mbox{\scriptsize pse}} \ +\ E_{\mbox{\scriptsize end}} \, .
\label{hbetax}
\eea
Here we have relabeled the remaining parameters as $\sigma$ and $\lambda$ in both
terms.  The three terms identified in the second relation correspond to the three terms in square brackets
of the first relation.   These terms involve scalar propagators only, and are represented by Fig.~\ref{fig:sespse}.
 We refer to the first term in brackets as the 3-scalar
integral, $E_{3s}$ (Fig.~\ref{fig:ses}), in which the end of one of the scalar propagators is fixed at the cusp by the
eikonal equation of motion. We will call the second term the
  ``pseudo-self-energy", $E_{pse}$ [Fig.~\ref{fig:pse}], since two scalar propagators form a
loop and attach to the Wilson line at the same point. Finally, the
third term, $E_{\rm end}$ [Fig.~\ref{fig:end}], in which
$\lambda_2=\Lambda$ for finite $\Lambda$ will be referred to as the
``end-point" diagram for this case. As noted above, the cusp itself is defined without the end-point diagram, but we will return to it in our discussion of Wilson line polygons below.

We can identify the sources of subdivergences in the expressions of Eq.\ (\ref{hbetax}) by finding points where the $x^\mu$ integral is pinched between coalescing singularities \cite{Erdogan:2013bga}.
In the 3-scalar term $E_{\rm 3s}$, the integration contours of the light cone component
$\beta_1\cdot x$ and two-dimensional transverse components $x_\perp$
are pinched when $x^\mu=\zeta \beta_1^\mu$, with
$0<\zeta<\sigma$, and also when $x^\mu =\eta \beta_2^\mu$, with
$0<\eta<\lambda$\@.   For fixed $\lambda$ and $\sigma$
these are the singular subdivergences referred to above, in which the
point $x^\mu$ approaches the path in the $\beta_1$ or $\beta_2$ directions, respectively.  In either case
two lines are forced to the light cone on one of the Wilson lines, while the third line may attach
anywhere on the opposite-moving line. There is no corresponding
pinch in the pseudo-self-energy term, and this diagram, along with the
self-energy diagrams, has only a single ultraviolet pole at fixed
$\lambda$ and $\sigma$, which is removed by the standard
renormalization of the gauge theory.

The integration of the 3-scalar term has been in the literature for a long time, but some details are given in the Appendix, to derive it as a coefficient times the scaleless integrals over parameters $\lambda$ and $\sigma$. We find 
\bea 
E_{3s} & = & \left(\frac{\alpha_s}{\pi}\right)^2
 C_AC_F\,  \frac{\Gamma(1-2\vep)\Gamma(1+\vep)\Gamma(1-\vep)}{16\,\vep^2}\,
 (2\pi\mu^2)^{2\vep} \int_0^\infty \frac{d\lambda\
   d\sigma}{\hspace{4mm} (\lambda\sigma)^{1-2\vep}} \ .
 \label{Fses}
\eea
We have taken the upper limits to infinity at this point, because we are
interested in the (unrenormalized) cusp integral.

The pseudo-self-energy term in Eq.\ (\ref{hbetax}) inherits the entire
ultraviolet divergence of the diagram $E_{\rm 3g}$,
Fig.~\ref{fig:3g} at fixed $\lambda$ and $\sigma$, and requires a counterterm that is part of the
web, rather than cusp, renormalization. The result is
\be
E_{\mbox{\scriptsize pse}}  = - \left(\frac{\alpha_s}{\pi}\right)^2
 C_AC_F\,\frac{1}{16\,\vep}\,
\int_0^\infty \frac{d\lambda\, d\sigma}{ \lambda\sigma} \left[ \frac{\Gamma^2(1-\vep)}{1-2\vep} (2\pi\mu^2\lambda\sigma)^{2\vep}
- \Gamma(1-\vep) (2\pi\mu^2\lambda\sigma)^{\vep}   \right]
\label{eq:Fpse}\, , 
\ee
with the same scaleless integral times a single-scale
constant. Finally, for the gluon self-energy diagrams,
Figs.~\ref{fig:se}--\ref{fig:se2}, we use the
renormalized one-loop gluon Green function in coordinate space. The
result for the self-energy contribution, $E_{\rm se}$
of Fig.~\ref{fig:se}, where the gluon connects both Wilson lines, can
be written as
\bea
E_{\mbox{\scriptsize se}} &\ =\ &
-\left(\frac{\as}{\pi}\right)^2 C_F \frac{1}{8\vep}  \int_0^\infty \frac{d\lambda\, d\sigma}{\lambda\sigma}\ 
\left[ \frac{\Gamma^2(1-\vep)}{1-2\vep} \left \{ \frac{(5-3\vep)C_A - 4 T_fn_f(1-\vep)}{3-2\vep} \  \right \}(2\pi\mu^2\lambda\sigma)^{2\vep}
\right. \nonumber\\
&\ & \left. \hspace{40mm}
-\  \Gamma(1-\vep)\, \left \{ \frac{5C_A - 4T_fn_f}{3} \right\}\,  (2\pi\mu^2\lambda\sigma)^{\vep}   \right]\
 +\ E_{\rm long} \, , 
 \label{eq:E-se}
\eea
where the (unrenormalized) longitudinal part of the Green function is given by 
\bea
E_{\rm long} &\ =\ &-  \left(\frac{\as}{\pi}\right)^2 C_F\,
\frac{\Gamma^2(1-\vep)}{32\,\vep^2\,(1+\vep)(1-2\vep)}\,\left \{
  \frac{(5-3\vep)C_A - 4 T_fn_f(1-\vep)}{3-2\vep}   \right \} \\
 & & \hspace{25mm}\times
  \int_0^\infty \ d\lambda\, d\sigma\ \frac{\partial}{\partial \lambda}\ \frac{\partial}{\partial \sigma}\
\Big[ (\pi\mu^2(\beta_2\lambda-\beta_1\sigma)^2)^{2\vep}
\Big]\, .  \nonumber
\eea
The function $E_{\rm long}$ comes from the coordinate-space transform
of the $q^\mu q^\nu$ term in the gluon self energy, and reduces to
total derivatives in both $\sigma$ and $\lambda$.   In momentum space,
the $q^\mu q^\nu$ terms decouple from the gauge-invariant cusp
algebraically in the sum over diagrams, assuming that the external
Wilson lines carry no momentum.   To define such derivative terms in
coordinate space for the cusp requires the introduction of small but
nonzero $\beta_1^2$ and $\beta_2^2$, and with this infrared
regularization, the longitudinal term above cancels the corresponding
term for the self-energy diagram of Fig.~\ref{fig:se2}, up to 
end-point contributions analogous to $E_{\rm end}$ in Eq.\ (\ref{hbetax}), which we have discarded in the calculation of the cusp contribution from $E_{3g}$ above.   We will once again neglect such terms for the purposes of this calculation, but will return to this question in the next subsection.

To check the finiteness and structure of the sum of these two-loop web
diagrams, we expand them in $\vep$, keeping all terms that can contribute
ultraviolet poles to the cusp. The (two) three-gluon diagrams plus the
crossed ladder gives 
\be
E_{\mbox{\scriptsize cross}}+2E_{3s} \  =\
\frac{1}{8}\left(\frac{\alpha_s}{\pi}\right)^2 C_FC_A \left (2\pi
  e^{\gamma_E}\mu^2\right)^{2\vep} 
\left(\frac{\pi^2}{3} + 2\vep\, \zeta_3 + {\cal O}(\vep^2) \right)  
\int_0^\infty \frac{d\lambda\ d\sigma}{\hspace{4mm}
  (\lambda\sigma)^{1-2\vep}}  \,  .
\label{eq:F3gexpand} 
\ee
Thus, as anticipated, the ultraviolet poles  from the subdivergences
of the web cancel, leaving only the overall scaleless integrals, whose
singular behavior can be associated with  hard, soft, and collinear
configurations for all of the lines of the web together. The $\pi^2$
term will contribute to $\Gamma_{\rm cusp}$ and the $\vep\zeta_3$ term to
$G_{\rm eik}$\@. We next expand the integrands of $E_{\rm se}$ and
$E_{\rm pse}$ at two loops, Eqs.~(\ref{eq:E-se}) and (\ref{eq:Fpse})
to order $\vep$, 
\bea
E_{\rm se}\ +\ 4E_{\rm pse} &=&
-\left(\frac{\as}{\pi}\right)^2 C_F \frac{1}{8}  \int_0^\infty \frac{d\lambda\, d\sigma}{\lambda\sigma}
\left[  \left \{ 1+ \vep^2\frac{\pi^2}{12}\right \} \frac{1}{\vep} b_0\, \left [(2\pi\mu^2e^{\gamma_E}\lambda\sigma)^{2\vep}  - (2\pi\mu^2e^{\gamma_E}\lambda\sigma)^{\vep} \right]  \right.
\nonumber \\
&\ &  \left. \hspace{-20mm} + \left  \{  \left( \frac{67}{9} 
\ C_A - \frac{20}{9} n_fT_f\right ) +
    \vep\left( \frac{404}{27}C_A-\frac{112}{27}n_fT_f +
      \frac{\pi^2}{12}b_0 \right ) \right \}
  (2\pi\mu^2e^{\gamma_E}\lambda\sigma)^{2\vep}  \right]\, . 
   \label{eq:E-se-plus-Fpse}
\eea
The terms proportional to $b_0/\vep$ serve to evolve the one-loop web, Eq.\ (\ref{eq:webLO}) to the scale $1/\lambda\sigma$ times constants.   

Combining Eqs.\ (\ref{eq:F3gexpand}) and (\ref{eq:E-se-plus-Fpse}), we find the explicit terms in the web expansion, Eq.\ (\ref{eq:w-vep-expand}).  In a scheme where logs of factors $2\pi e^{\gamma_E}$  are absorbed into the definition of $\as(1/\lambda\sigma)$, we have for the terms in Eq.\ (\ref{eq:w-vep-expand}),
\bea
w_0(\as)\ &=& \ -\  \frac{\as}{2\pi}\ C_F   \ -\  \left(
  \frac{\as}{\pi} \right )^2\,  \frac{C_F}{2}\, \left (
  \left[\frac{67}{9} -\frac{\pi^2}{3}\right]C_A - \frac{20}{9} n_fT_f
\right ) \ +\ \dots \, ,
\nonumber\\
w_1(\as)\ &=&\ -\  \left(\frac{\alpha_s}{\pi}\right)^2\ \frac{C_F}{8}\
\left ( \left [\frac{404}{27}\ -\ 2\zeta_3 \right ]C_A\ -\
  \frac{112}{27}n_fT_f\ +\ \frac{\zeta_2}{2}\, b_0\right) \ +\ \dots
\, ,
\nonumber\\
w_2(\as)\ &=&\  -\ \frac{\as}{2\pi}\ C_F\, \frac{\pi^2}{12} \ +\ \dots 
\label{eq:F3g-result}\, ,
\eea
where omitted terms are higher order in $\as$ or do not contribute to
the cusp ultraviolet poles.   The term linear in $\vep$ begins at
order $\as^2$, but the single pole also gets a contribution from the
$\vep^2$ term at one loop, when combined with the running of the
coupling. With these results in hand, we can return to Eq.\ (\ref{eq:w-vep-expand}) and expand $\as(1/\lambda\sigma)$ in terms of the coupling at a fixed scale, $\as(\mu^2)$ using (\ref{eq:alpha-ls-mu}).  This enables us to derive the single ultraviolet pole in  $E$ to order $\as^2$, and hence the  anomalous dimension $G_{\rm eik}$ at two loops,
\be
G_{\rm eik}
\ =\ \frac{1}{2}\ C_FC_A\  \left( \frac{\as}{\pi} \right)^2\ \left[ \left \{\frac{101}{27}\ - \frac{11}{72}\pi^2 \ -\ \frac{1}{2}\, \zeta_3\right \}C_A\ +\ \left\{ \frac{28}{27}\ - \frac{\pi^2}{18} \right\}n_fT_f\ \right] \, .
\label{eq:G-eik-2loop}
\ee
In Sec.\ \ref{sec:poly}, we will see the close relation of this result to
the ``collinear anomalous dimension'' derived long ago in
Ref.~\cite{Korchemskaya:1992je} for a closed polygon of Wilson lines
of finite size.

\subsection{Web integrals, end points and gauge invariance}

   A self-contained coordinate-space derivation of Eq.~(\ref{eq:web2}),
generalizing the renormalization analysis of Ref.~\cite{Brandt:1981kf}
for massive Wilson lines is given in~\cite{Erdogan:2014gha}\@. 
Here, however, we will generalize our
prescription for the calculation of the gauge-invariant cusp anomalous
dimension. As we have seen, this requires us to find in coordinate
space the analog of the action of momentum-space Ward identities that
ensure the gauge invariance of the \mbox{S-matrix}~\cite{'tHooft:1971fh}\@.

In the following brief but all-orders discussion
we follow Ref.~\cite{Mitov:2010rp}
and write the exponent as a sum over the numbers, $e_a$, of gluons attached 
to the two Wilson lines, 
of velocity
$\beta_a$, $a=1,2$\@.    We note, however, that the argument extends to any number of lines.
The web diagrams are integrals over the
positions $\lambda_j\beta_1$ and $\sigma_k\beta_2$ of
these ordered vertices of a function ${\cal
  W}_{e_1,e_2}\left(\{\lambda_j\},\{\sigma_k\}\right)$, which includes
the integrals over all the internal vertices of the corresponding web diagrams.
In the notation of Ref.~\cite{Mitov:2010rp} we then have
 at $n$th order $(n\ge e_1+e_2)$,
\be
E^{(n)} =
 \sum_{e_2=1}^{n-1}\  \sum_{e_1=1}^{n-e_2} \
\prod_{j=1}^{e_1}\ \int_{\lambda_{j-1}}^\infty\! d\lambda_j\
\prod_{k=1}^{e_2}\ \int_{\sigma_{k-1}}^\infty\! d\sigma_k\
{\cal W}_{e_1,e_2}^{(n)}\left(\{\lambda_j\},\{\sigma_k\}\right)\, ,
\label{eq:calFcalW}
\ee
with $\lambda_0,\sigma_0\equiv 0$\@. Here  we expand
functions as $E=\sum (\alpha_s/\pi)^nE^{(n)}$\@.  
We can use the notation of Eq.\ (\ref{eq:calFcalW}) to generalize our treatment of the three-gluon diagram and self-energy diagrams above.   
First, we isolate those contributions to ${\cal W}_{e_1,e_2}^{(n)}\left(\{\lambda_j\},\{\sigma_k\}\right)$ that are of the form of total derivatives in the largest path parameters, $\lambda_{e_1}$, $\sigma_{e_2}$, and whose upper limits vanish when the end points of ordered exponentials are taken to infinity for fixed values of the internal vertices of the web. We represent this separation as, 
\bea
{\cal W}_{e_1,e_2}^{(n)}\left(\{\lambda_j\},\{\sigma_k\}\right)
&=&
\ \frac{\partial}{\partial\, \lambda_{e_1}}\ {\cal
  X}^{(\lambda)(n)}_{e_1,e_2}\left(\{\lambda_j\},\{\sigma_k\}\right)\ 
+\ \frac{\partial}{\partial\, \sigma_{e_2}}\ {\cal
  X}^{(\sigma)(n)}_{e_1,e_2}\left(\{\lambda_j\},\{\sigma_k\}\right) \\
&\ & 
+\ \frac{\partial}{\partial\, \lambda_{e_1}}\ \frac{\partial}{\partial\, \sigma_{e_2}}\ {\cal X}^{(\lambda\sigma)(n)}_{e_1,e_2} \left(\{\lambda_j\},\{\sigma_k\}\right)
\  +\  
\overline{\cal
  W}_{e_1,e_2}^{(n)}\left(\{\lambda_j\},\{\sigma_k\}\right)\, , \nonumber
\eea
where the ${\cal X}^{(I)}$, $I=\lambda,\, \sigma\, , \lambda\sigma$,
are functions whose derivatives are taken by $\lambda_{e_1}$,
$\sigma_{e_2}$ or both, and which vanish when $\lambda_{e_1}$ and/or $\sigma_{e_2}$ are taken to infinity with other integration variables held fixed.
The function $\overline{W}$ is the remaining web
integrand. To determine the cusp, we evaluate the total derivatives at the lower limits, $\lambda_{e_1}=\lambda_{e_1-1}$,
$\sigma_{e_2}=\sigma_{e_2-1}$ or both, discarding the upper limits, as $E_{\rm end}$ in the two-loop case above.   We then relabel the largest remaining $\lambda_j$ integral (either $\lambda_{e_1}$ or $\lambda_{e_1-1}$) as $\lambda$, and integrate over the rest of the $\lambda_j$, up to $\lambda$.   The $\sigma_k$ parameters are treated in just the same way.   In this manner, we find for the web function in Eq.\ (\ref{eq:web2}), the form
\bea
w \left (\as(1/\lambda\sigma,\vep),\lambda\sigma\mu^2,\vep\right)
&=&
 \sum_{e_2=1}^{n-1}\  \sum_{e_1=1}^{n-e_2} \
\prod_{j=1}^{e_1}\ \int_{\lambda_{j-1}}^\lambda\! d\lambda_j\ 
\prod_{k=1}^{e_2}\ \int_{\sigma_{k-1}}^\sigma\! d\sigma_k\ \delta(\lambda_{e_1}-\lambda)\, \delta(\sigma_{e_2}-\sigma)
\nonumber\\
&\ & \hspace{-40mm} \times\ \Bigg[
-\  \delta(\lambda_{e_1-1}-\lambda)\ {\cal X}^{(\lambda)(n)}_{e_1,e_2}\left(\{\lambda_j\},\{\sigma_k\}\right)
-\   \delta(\sigma_{e_2-1}-\sigma)\ {\cal X}^{(\sigma)(n)}_{e_1,e_2}
\left(\{\lambda_j\},\{\sigma_k\} \right) \\
&\ & \hspace{-32mm}
+\  \delta(\lambda_{e_1-1}-\lambda)\,   \delta(\sigma_{e_2-1}-\sigma)\  {\cal X}^{(\lambda\sigma)(n)}_{e_1,e_2} \left(\{\lambda_j\},\{\sigma_k\}\right)
\ +\ 
 \overline{\cal
   W}_{e_1,e_2}^{(n)}\left(\{\lambda_j\},\{\sigma_k\}\right)\,
 \Bigg]\, . \nonumber
\eea
Once web diagrams are summed over at any order, this form is gauge
invariant, and produces the same cusp integrand for finite lines as
for infinite lines.   This is because the infinitesimal gauge
variation of a product of Wilson lines as in Eq.\ (\ref{eq:Gammadef})
produces a ghost propagator ending on the ends of the lines, which
vanishes when those lines are taken to infinity \cite{'tHooft:1971fh}.
Even if the ends of the lines are at finite distances, the
prescription to discard the upper limit of total derivatives
automatically removes these gauge variations.   When the end points,
which generalize $E_{\rm end}$ in Eq.\ (\ref{hbetax}) in our
discussion above, are at finite distances, however, we must keep these
terms and combine them with the remainder of the diagrams of the graph
to derive the full, gauge-invariant result. 

\section{Applications to Polygon Loops}
\label{sec:poly}

The above reasoning leads to a number of interesting results for 
polygonal closed Wilson loops \cite{Alday:2007hr,Alday:2008yw,Drummond:2007aua}.
These amplitudes also exponentiate in perturbation theory in terms of
webs~\cite{Drummond:2007aua}\@. To this observation we may apply once
again the lack of subdivergences for webs.   

\begin{figure}[b]
\centering 
\includegraphics[scale=1]{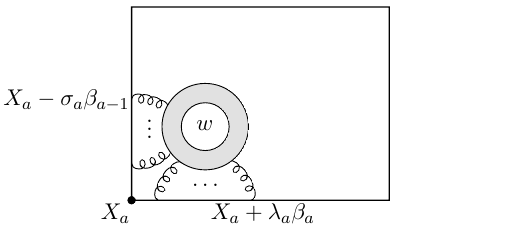}
\caption{A single-cusp web $W_a$, in the sum
of Eq.\ (\ref{eq:Lfsum1}).}
\label{fig:cuspweb}
\end{figure}

\begin{figure}[b]
\centering
\subfigure[]{\includegraphics[scale=1]{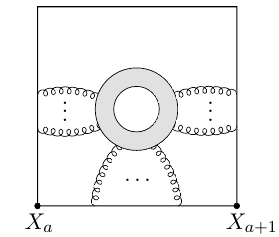}\label{fig:sw1}}
\quad
\subfigure[]{\includegraphics[scale=1]{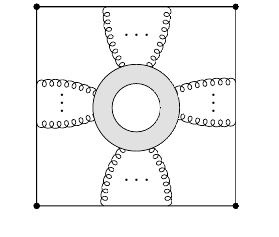}\label{fig:sw2}}

\caption{\subref{fig:sw1}~A ``side'' web $W_{a+1,a}$ in Eq.~(\ref{eq:Lfsum1}), in this case associated
with the lightlike side between $X_{a}$ and $X_{a+1}$\@. \subref{fig:sw2}~A web
that contributes to $W_{\rm plane}$ in Eq.\ (\ref{eq:Lfsum1}).}

\label{fig:sidewebs}

\end{figure}

Generic diagrams for quadrilateral loops are shown in
Figs.~\ref{fig:cuspweb} and \ref{fig:sidewebs}. 
In Fig.~\ref{fig:cuspweb}, for example, the $a$th vertex of the polygon
represents a cusp vertex  that connects two Wilson 
lines, of velocity $\beta_{a-1}$ and $\beta_a$, with $\beta_0\equiv \beta_4$.   

Exponentiation in coordinate space implies that the logarithm of a
polygon~$P$ is a sum of the web configurations illustrated by the figures,
\bea
\ln P &=& \sum_{{\rm cusps}\ a} W_a \ 
+ \sum_{{\rm sides}\ \{a+1,a\}}W_{a+1,a}\ 
+\ W_{\rm plane}\, .
\label{eq:Lfsum1}
\eea
The  first terms organize webs associated entirely with one of the cusps of the
polygon, constructed in terms of the
coordinate webs identified above.   Because each edge is of finite length, we must now retain the additional
gauge-variant terms associated with the end-point contributions ($E_{\rm end}$ above),
which are to be combined with gauge-variant end points from webs connecting three or four sides to derive a gauge-invariant result.   
The cancellation of subdivergences in webs  implies that after a sum over diagrams,
 only  the cusp poles and  a single, overall
collinear singularity survives  \cite{Drummond:2007aua,Erdogan:2014gha}.  
There remains a finite contribution from webs that connect
all four (or in general more) of the Wilson lines, and these are represented by the final term in (\ref{eq:Lfsum1}).

Evidently, the single-cusp contribution,
$W_a(\beta_{a},\beta_{a-1})$ has the same gauge-invariant integrand as for the finite Wilson lines
in Eq.\ (\ref{eq:web3}), in terms of the lengths $L_a$ of the sides of the polygon,
between vertices $a$ and $a+1$
\be
W_{a}(\beta_a,\beta_{a-1},L_a,L_{a-1})
\ =\ 
\int_{0}^{L_a} \! \frac{d\lambda_a}{\lambda_a}\, 
\int^0_{-L_{a-1}}\! \frac{d\sigma_a}{\sigma_a}\, w(\alpha_s(1/\lambda_a\sigma_a,\vep),\varepsilon)\, .
\label{eq:web_sides}
\ee
The web function $w$ for the cusp  can depend only on the scalar products
of the velocities, and we may assume for simplicity that these are all of the same order.

The two-cusp contributions connect three sides, and the only available singular configuration is
when all lines in the web are parallel to the side between the two adjacent vertices.   
 The only invariants on which the web can then depend are of the form $L_a\eta$, with
 $L_a$ the length of this side, and $\eta$ a typical distance of vertices in the web from the side.
 As a result, the general form of the $W_{a+1,a}$ in Eq.\ (\ref{eq:Lfsum1}) is
\bea
W_{a+1,a}(L_a)\ =\ \int_0^{L_{a}} \frac{d\eta}{\eta}\ w_{a+1,a}\left(
  \alpha_s(\eta L_{a},\vep)\right) \, ,
\eea
for a function $w_{a+1,a}(\alpha_s)$\@, where we assume all the sides are of a similar length. Finally, for the diagrams in
which the web is stretched out between more than three sides of a 
polygon (in this case, the web is connected to all four sides of the
quadrilateral), $W_{\rm plane}$, the only scale available is the area of the quadrilateral, and these web contributions are an expansion in the coupling evaluated at the inverse area,
with finite coefficients.

The two-loop diagrams for all of these topologies were computed in~\cite{Drummond:2007aua}\@.   
We note that in the results quoted there, the cusp anomalous dimension does not appear until
all diagrams of the topologies of $W_a$ and $W_{a+1,a}$ are combined.   Following the prescription for
the web integrand given above, however, the two-loop cusp is associated entirely with the
diagrams dressing a single corner, $W_a$, precisely because the gauge-variant end-point contributions $E_{\rm end}$
of Eq.\ (\ref{hbetax}) are not included in that object.  
For polygons, these gauge-variant terms at two loops, or any
order, cancel contributions from the two-cusp contributions
$W_{a+1,a}$, which also give rise to gauge-variant terms that
cancel those from planar diagrams. These gauge-variant terms
contain subdivergences in general. The complete result, of course, is
gauge invariant and corresponds at two loops to the full calculation
in Refs.~\cite{Korchemskaya:1992je} 
and \cite{Drummond:2007aua}\@.

For polygons, the renormalization group equation has been given in
\cite{Korchemskaya:1992je}, 
\bea
\frac{d}{d\, \ln\mu^2} \ P_{\rm ren}  \ =\ -\frac{1}{2}\; \sum_a \Gamma_{\rm cusp}(\alpha_s(\mu^2))\ \ln(\mu^2L_aL_{a-1}\beta_a\cdot \beta_{a-1}) \ -\ \Gamma_{\rm co}\left(\alpha_s(\mu^2)\right)\, ,
\eea
where the $L_a$ and $\mu$-dependence of the first term is
characteristic of cusps with lightlike Wilson
lines~\cite{Korchemsky:1987wg}, and where the second term, $\Gamma_{\rm
  co}$ was called the collinear anomalous dimension in
Ref~\cite{Korchemskaya:1992je}. Aside from overall factors associated
with the number of sides of the polygon, the collinear anomalous
dimension for the quadrilateral is identical to $G_{\rm eik}$ in
Eq.~(\ref{eq:G-eik-2loop}), except for the coefficient of $\zeta_3$, 
which differs due to extra diagrams that connect three sides of the
quadrilateral.

Polygons of this sort have been studied
in the context of a duality to scattering amplitudes in conformal
theories~\cite{Alday:2007hr,Drummond:2007aua}\@. 
Here,
we consider a four-sided polygon 
that projects to a square in the $x_1/x_2$ plane, with side $X$,
as in Figs.~\ref{fig:cuspweb}--\ref{fig:sidewebs}\@. In four dimensions, the loop starts at the origin, travels along the
plus-$x_1$ direction for a ``time'' $X^0=X$, then changes direction
to $x_2$ for time $X$, and then moves backwards in time and space,
first in the $x_1$ direction, then $x_2$, back to the origin.
We can now use the coordinates $x_1$ and $x_2$ to 
define parameters $\lambda_a$ and $\sigma_a$ for
each of the cusp integrals $W_a$ in Eq.~(\ref{eq:web_sides}),
\be\begin{array}{lcl}
\sigma_1 = -x_2\, ,& \qquad &\lambda_1 = x_1\, , \\
\sigma_2 = x_1-X\, ,& \qquad & \lambda_2 = x_2\, , \\
\sigma_3 = x_2-X\, , & \quad & \lambda_3 = X-x_1\, , \\
\sigma_4 = -x_1\, , & \quad & \lambda_4 = X-x_2\, .
\end{array}\ee
In 
this notation,
we can add the four cusp
web integrals of Eq.~(\ref{eq:web_sides}), 
to get
a single
integral over $x_1$ and $x_2$.   The web functions, of course,
depend on the particular forms of $\lambda$ and $\sigma$
above. We find
\be
\sum_{a=1}^4 W_a(\beta_a,\beta_{a-1})
 \,= \, \int_0^X\!\! dx_1\int_0^X\!\! dx_2\,
\frac{(X-x_2)[(X-x_1) w_1 + x_1 w_2]
+
x_2[x_1 w_3 + (X-x_1) w_4]}
{x_1(X-x_1)x_2(X-x_2)}
\, ,
\label{eq:sum_webs}
\ee
where $w_a\equiv w(\alpha_s(\lambda_a(x_1,x_2)\sigma_a(x_1,x_2)))$.
For a conformal theory, all dependence on the $\sigma_a$ and $\lambda_a$
is in the denominators and we can sum over $a$ to get a result in terms
of a constant web function $w_0$. Changing variables to $y_a=1-2x_a/X$, we
derive the unregularized form found from the analysis of extremal
two-dimensional surfaces embedded in a five-dimensional background in~\cite{Alday:2007hr}, 
\be
\sum_{a=1}^4 W_a(\beta_a,\beta_{a-1})
\ =\ \int_{-1}^1 dy_1\, \int_{-1}^1 dy_2\,
\frac{4 w_0}
{(1-y_1^2)(1-y_2^2)}
\, ,
\label{eq:sum_conformal_webs}
\ee
to which we should add
the collinear and finite multi-cusp contributions of Fig.~\ref{fig:sidewebs}.

\section{Conclusions}

We have found that when the massless cusp is analyzed in coordinate space,
it is naturally written as the exponential of a two-dimensional integral.   The integrand, 
a web function,
depends on the single invariant scale
through the running of the coupling, which for a theory that is conformal
in four dimensions agrees with 
strong-coupling results~\cite{Alday:2007hr,Alday:2008yw,Kruczenski:2002fb}\@.
This agreement extends to aspects of closed, polygonal Wilson loops.
These results do not rely on a planar limit \cite{'tHooft:1973jz},
but it is natural to conjecture that for large $N_c$ the 
integral may take on an even more direct interpretation in terms of
surfaces for nonconformal theories.

In QCD, of course, our 
explicit
knowledge of the web function is limited
to the first few terms in the perturbative series, which run out of predictive power
as the invariant distance increases. The integral forms derived above, however,
hold to all orders in perturbation theory, and may point to an
interpolation
between short and long distances.

\acknowledgments
We thank G. P. Korchemsky and B. van Rees for helpful discussions. This work was
supported by the National Science Foundation, 
Grants No. PHY-0969739 and No. PHY-1316617.

\begin{appendix}

\section{Two-loop Integrals}

\subsection{The 3-scalar integral}

To evaluate the the 3-scalar term in Eq.~(\ref{hbetax}), we integrate over the position of the three-gluon vertex
after combining the denominators by Feynman parametrization. Introducing the
Feynman parameters $\alpha_1$ and $\alpha_2$, the 3-scalar
contribution is given by
\be\begin{split} E_{3s}\ =\ & -\
  \mathcal{N}_{3g}(\vep)\int_0^\infty d\lambda\,d\sigma\int
  d^{4-2\vep}y\:\frac{\Gamma(3-3\vep)}{\Gamma^3(1-\vep)}  \\
 &\quad\times\int^1_0d\alpha_1\int^{1-\alpha_1}_0\!\!d\alpha_2\frac{(1-\alpha_1-\alpha_2)^{-\vep}\alpha^{-\vep}_1\alpha^{-\vep}_2}{\left[-y^2+2\alpha_2(1-\alpha_1-\alpha_2)\lambda\sigma+i\epsilon\right]^{3-3\vep}}
\ , \end{split}\ee
where $y\equiv
x-\alpha_2\lambda\beta_1-(1-\alpha_1-\alpha_2)\sigma\beta_2$\@. The
integral over $y$ is straightforward after doing a clockwise Wick rotation,
\be\begin{split}
E_{3s}\ = \  &-\ \mathcal{N}_{3g}(\vep)\left(
  \frac{-i\pi^{2-\vep}}{2^{1-2\vep}}\frac{\Gamma(1-2\vep)}{\Gamma^3(1-\vep)}\right)\int_0^\infty \frac{d\lambda\
   d\sigma}{\hspace{4mm} (\lambda\sigma)^{1-2\vep}} \\
&\quad\times \int^1_0d\alpha_1\int^{1-\alpha_1}_0d\alpha_2\,(1-\alpha_1-\alpha_2)^{-1+\vep}\,\alpha^{-\vep}_1\,\alpha^{-1+\vep}_2 
 \ . \end{split}
 \label{eq:E3s-app}
 \ee
The integrals over Feynman parameters
$\alpha_1,\alpha_2$ now factor  from the integrals over eikonal
parameters $\lambda,\sigma$\@. After a change of variables
\mbox{$\eta\equiv\alpha_2/(1-\alpha_1)$}, they can be integrated independently, 
\be
\int^1_0d\alpha_1\,\alpha^{-\vep}_1(1-\alpha_1)^{2\vep-1}\int^{1}_0d\eta\,\eta^{\vep-1}(1-\eta)^{\vep-1}\ 
=\ \frac{1}{\vep^2}\Gamma(1-\vep)\Gamma(1+\vep) \ . \ee
In Eq.\ (\ref{eq:E3s-app}), this gives the scaleless $\lambda,\,
\sigma$ integral times a constant with a double pole in $\vep$, 
given in Eq.~(\ref{Fses})\@.

\subsection{The end-point term}

We now return to the $\lambda_2=\Lambda$ end-point contribution from
the second term on the right-hand side of Eq.\ (\ref{3v01}), which
vanishes in the $\Lambda\rightarrow \infty$ limit for any fixed values
of the vertex~$x^\mu$\@. If we integrate over $x^\mu$ first, however, we
get a singular contribution, associated with the renormalization of a
Wilson line of finite length. 
It cancels in the gauge-invariant  polygons discussed in Sec.\
\ref{sec:poly}, and extensively in
Refs.~\cite{Korchemskaya:1992je,Drummond:2007aua}\@. After the $x^\mu$
integral, we have
\bea
E_{\mbox{\scriptsize end}} &\ =\ & 
 -\ \mathcal{N}_{3g}(\vep)\left(\frac{-i\pi^{2-\vep}}{2^{1-2\vep}}\frac{\Gamma(1-2\vep)}{\Gamma^3(1-\vep)}\right)
  \int_0^\Sigma \frac{ d\sigma}{\sigma^{1-2\vep}}  \int_0^\Lambda d\lambda \\
&\ &\quad\times\int^1_0d\alpha_1\int^{1-\alpha_1}_0\!\!d\alpha_2\,
 \alpha_1^{\vep-1}(1-\alpha_1-\alpha_2)^{-\vep}\alpha_2^{-\vep}
 \left[\alpha_2\Lambda + (1-\alpha_1-\alpha_2) \lambda \right]^{-1+2\vep}
\ . \nonumber 
\eea
Changing variables to $\eta=\alpha_2/(1-\alpha_1)$, we find a form that is easy to evaluate,
\bea
E_{\mbox{\scriptsize end}} &\ =\ & -\
  \mathcal{N}_{3g}(\vep)\left(\frac{-i\pi^{2-\vep}}{2^{1-2\vep}}\frac{\Gamma(1-2\vep)}{\Gamma^3(1-\vep)}\right)
    \int_0^\Sigma \frac{ d\sigma}{\sigma^{1-2\vep}}  \label{eq:Eend-result} \\
 &\ & \hspace{5mm} \times\ \int^1_0d\alpha_1\,\alpha_1^{\vep-1}
 \int^1_0\!\!d\eta\, (1-\eta)^{-\vep}\eta^{-\vep}\ 
  \int_0^\Lambda\ d\lambda 
 \left[\eta\Lambda + (1-\eta) \lambda \right]^{-1+2\vep}
\nonumber \\
 &\ =\ &  \left(\frac{\alpha_s}{\pi}\right)^2\ C_AC_F\ \left(2\pi \mu^2\Lambda\Sigma\right)^{2\vep}\ \frac{1}{64\vep^4}\
 \left[ \Gamma(1-2\vep)\, \Gamma(1-\vep)\, \Gamma(1+\vep) \ -\
   \Gamma^2(1-\vep) \right]\, . \nonumber 
\eea
If we add this result to the expressions found by integrating the
$\sigma$ and $\lambda$ integrals of $E_{3s}$, Eq.~(\ref{Fses}) and
$E_{pse}$, Eq.~(\ref{eq:Fpse}), over the finite intervals of $0$ to
$\Sigma$ and $\Lambda$, we recover the expression quoted for this
diagram in Refs.~\cite{Korchemskaya:1992je,Drummond:2007aua}\@.

\end{appendix}

\end{document}